\documentclass[prl,twocolumn,preprintnumbers,amsmath,amssymb,superscriptaddress,floatfix]{revtex4}

\usepackage{graphicx}
\usepackage{dcolumn}
\usepackage{bm}

\begin{document}

\title{Detection, spectroscopy and state preparation of a\\ single praseodymium ion in a crystal}

\author{Tobias Utikal}
\affiliation{Max Planck Institute for the Science of Light and Friedrich-Alexander-Universit\"at Erlangen-N\"urnberg (FAU),
D-91058 Erlangen, Germany}
\author{Emanuel Eichhammer}
\affiliation{Max Planck Institute for the Science of Light and Friedrich-Alexander-Universit\"at Erlangen-N\"urnberg (FAU),
D-91058 Erlangen, Germany}
\author{Lutz Petersen}
\affiliation{Current address: OSRAM GmbH, Berliner Allee 65, 86153 Augsburg, Germany}
\author{Alois Renn}
\affiliation{Laboratory of Physical Chemistry, ETH Zurich, 8093 Zurich, Switzerland}
\author{Stephan G\"otzinger}
\affiliation{Max Planck Institute for the Science of Light and Friedrich-Alexander-Universit\"at Erlangen-N\"urnberg (FAU),
D-91058 Erlangen, Germany}
\author{Vahid Sandoghdar}
\affiliation{Max Planck Institute for the Science of Light and Friedrich-Alexander-Universit\"at Erlangen-N\"urnberg (FAU),
D-91058 Erlangen, Germany}


\begin{abstract}

Solid-state emitters with atom-like optical and magnetic transitions are highly desirable for efficient and scalable quantum state engineering and information processing. Quantum dots, color centers and impurities embedded in inorganic hosts have attracted a great deal of attention in this context, but influences from the matrix continue to pose challenges on the degree of attainable coherence in each system. We report on a new solid-state platform based on the optical detection of single praseodymium ions via 4f intrashell transitions, which are well shielded from their surroundings. By combining cryogenic high-resolution laser spectroscopy with fluorescence microscopy, we were able to spectrally select and spatially resolve individual ions. In addition to elaborating on the essential experimental steps for achieving this long-sought goal, we demonstrate state preparation and read out of the three ground-state hyperfine levels, which are known to have lifetimes of the order of hundred seconds. 

\end{abstract}

\maketitle

One of the exciting scientific challenges of modern times is the realization of quantum networks that process information in unconventional nonclassical ways~\cite{kimble08quantum}. This would entail a sophisticated mesh of nodes for efficient storage of quantum information, its processing and read out as well as mechanisms for connecting the individual units in a low-loss and efficient manner. The success of this ambitious plan relies on the availability of suitable materials, much in the same way that the properties of silicon have played a key role in the rapid advances of microelectronics. A particularly elegant solution would be to combine narrow optical transitions with long coherence times of electron and nuclear spins~\cite{Awschalom:13} in a solid-state system that lends itself to on-chip photonics implementations~\cite{OBrien:07}. While no individual platform has been able to satisfy all these requirements, many pioneering efforts have made remarkable progress by investigating elementary processes in a wide range of systems such as atoms and ions in the gas phase, molecules, semiconductor quantum dots, color centers, and superconducting circuits. Here, we report on a new approach based on the direct all-optical detection of a single praseodymium ion by spectrally selecting it from an inhomogeneously broadened ensemble~\cite{Moerner:89} in yttrium orthosilicate (Y$_2$SiO$_5$, also called YSO). 

\begin{figure*}
\includegraphics[width=0.8\textwidth]{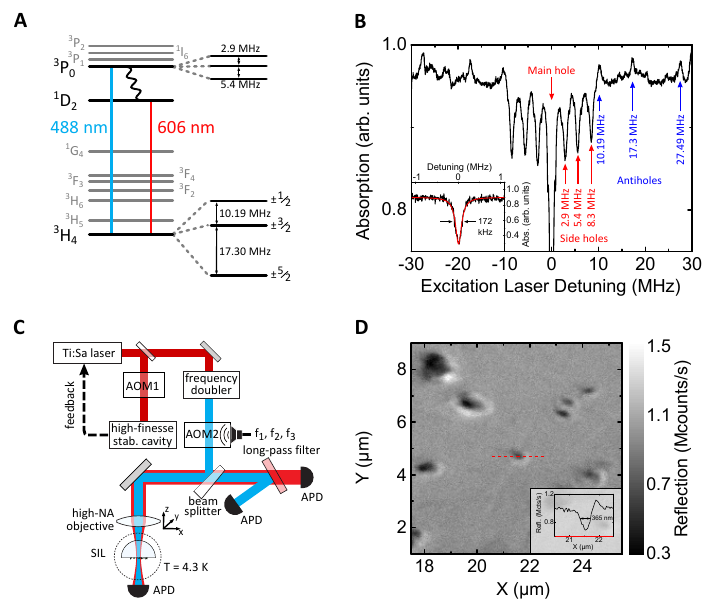}
\caption{\label{setup}
\textbf{A}, Level scheme of Pr$^{3+}$ in YSO. The ground  ($^{3}H_4$) and excited states ($^{3}P_0$) show hyperfine splittings into three doubly degenerate levels. The hyperfine level splittings of $^{3}P_0$ were measured for the first time in this report. \textbf{B}, Holeburning spectrum measured on an ensemble of Pr$^{3+}$ in a bulk YSO crystal. Side holes and anti-holes mark the excited and ground state hyperfine splittings, respectively. The inset displays a zoom of the main hole at low fluence excitation. \textbf{C}, Schematic of the experimental setup. A frequency-stabilized Ti:sapphire laser is frequency-doubled and used for excitation. An acousto-optical modulator (AOM1) scans the frequency of the fundamental light. A second modulator (AOM2) produces the three frequencies $f_1$, $f_2$ and $f_3$ described in the main text. Pr:YSO microcrystals are placed on the flat side of a hemispherical solid immersion lens (SIL) inside a liquid helium flow cryostat. The 488~nm excitation light is focused onto the microcrystals by a high-NA objective mounted on a 3D piezo stage. Fluorescence and the reflected light are detected through the same objective. \textbf{D}, Reflection scan image of the sample, visualizing the individual micro- and nanocrystals. The inset shows a cross section of the nanocrystal used in our measurements.}
\end{figure*}

A special feature of rare earth ions is that their 4f-shell transitions are shielded from their surroundings by closed outer shells, allowing for long coherence times~\cite{thiel11rare-earth-doped}. Figure~\ref{setup}A sketches some of the energy levels of Pr$^{3+}$ in the visible spectrum. Level $^3H_4$ acts as the electronic ground state with three doubly-degenerate hyperfine sublevels at frequency spacings of 10.19 and 17.3 MHz~\cite{equall95homogeneous}. Hole burning measurements have shown the lifetimes of these levels to be as long as 100~s~\cite{holliday93spectral, nilsson04hole-burning}, and spin coherence times up to one minute have been recently reported~\cite{Fraval:05, lovric11hyperfine, Heinze13minutememory}. The well-defined narrow optical transitions, long hyperfine coherence times and access to $\Lambda$ level schemes make rare earth ions highly attractive for quantum optics~\cite{longdell05stopped, Riedmatten:08, Heinze13minutememory}. On the other hand, the long excited-state lifetimes and the resulting weak emission have restricted previous studies to ensembles. Indeed, faint fluorescence, photobleaching and irreproducibility seem to have hampered progress in a few isolated attempts that have been reported over more than two decades~\cite{lange88observation, Yen89, Rodrigues-Herzog:00, bartko02observation, malyukin03single-ion}. A year ago, Kolesov \textit{et al.} presented a solid proof of detecting single Pr$^{3+}$ ions at room temperature~\cite{kolesov12optical}. These authors considered the direct detection of single ions via 4f-4f transitions to be unfeasible and circumvented the long optical coherence times by exciting the ion to the 5d band in a multiphoton process. The room-temperature operation of this pioneering work, however, does not allow addressing the narrow optical transitions and hyperfine levels of the ground state. Another very recent publication reports on electrical detection of erbium ions in a silicon nanotransistor structure at cryogenic temperatures~\cite{yin13optical}, but optical read out and integration in photonics networks were not accessible. 

In our laboratory we chose to excite Pr$^{3+}$ to the $^3P_0$ state because its lifetime of $1.95~\mu s$ is shorter than $166~\mu s$ for the $^1D_2$ state, which has been used in the great majority of the published research~\cite{nilsson04hole-burning, fraval04method, Fraval:05, longdell05stopped, lovric11hyperfine, Heinze13minutememory}. The correspondingly broader natural linewidth of 82 kHz instead of 1 kHz sets a less stringent requirement for the bandwidth of the laser to be used. When an ion is excited to the $^3P_0$ state, it can decay into each of the three ground state hyperfine levels via different pathways. The main channel involves a non-radiative relaxation to the $^1D_2$ state followed by fluorescence decay at a wavelength of 606 nm. Figure S1A in the Supplementary Materials provides more details on the emission spectrum.  In our work, we used long-pass filters to collect the fluorescence at wavelengths above 595 nm. 

Considering the long lifetime of the ground state hyperfine levels, excitation at only one laser frequency would quickly transfer the population of any of the hyperfine levels into the other two. The resulting population trapping leads to the inhibition of fluorescence and is the mechanism that allows spectral hole burning~\cite{holliday93spectral, nilsson04hole-burning}. In Fig.~\ref{setup}B we display an example of a hole burning spectrum obtained from a bulk crystal at T=4.3 K. Here, a strong narrow-band laser beam at a fixed frequency was used to optically pump the population from one of the $^3H_4$ hyperfine levels of a subclass of ions, thus ``burning" a spectral hole in the absorption profile of the sample. A weak probe laser was subsequently scanned over the spectral hole to report on the homogeneous broadening and the separation of the hyperfine levels. The spectral dips on the sides of the central hole in Fig.~\ref{setup}B reveal the hyperfine splittings of the $^3P_0$ state at 2.9 MHz and 5.4 MHz, measured for the first time in Pr:YSO. The anti-holes appear at the expected frequency spacings of 10.19 MHz and 17.3 MHz for the hyperfine levels of the ground state~\cite{equall95homogeneous}. The observed spectral features with linewidths of about 172~\textrm{kHz} in low-fluence hole burning experiments (see inset of Fig.~\ref{setup}B) are consistent with a homogeneous width of 82 kHz and a laser linewidth below 10~kHz (see Supplementary Materials). Repeated scans verify that the hole depth and spectral position persist in the bulk sample for minutes, pointing to the absence of noticeable spectral diffusion in this system. 

\begin{figure*}
\includegraphics[width=0.8\textwidth]{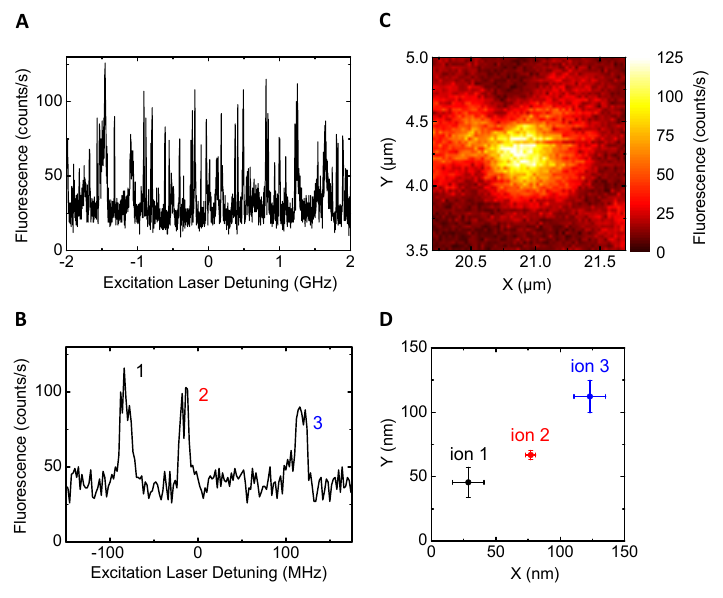}
\caption{\label{single_ion}
\textbf{A}, Fluorescence excitation spectrum recorded over several GHz. Narrow resonances are attributed to single ions. \textbf{B}, An excitation fluorescence spectrum recorded over about 300 MHz, consisting of three ions. \textbf{C}, Laser scanning fluorescence image of a single ion recorded on resonance. \textbf{D}, Super-resolution co-localization image of the three ions detected in (B).}
\end{figure*}

To detect single ions on a low background, it is desirable to reduce the observation volume and concentration by as much as possible. In our current work, we were restricted to commercially available Pr:YSO crystals with minimum doping of 0.005\%. Therefore, we milled a small piece of such a crystal to obtain micro- and nanocrystals. As sketched in Fig.~\ref{setup}C, the crystallites were deposited on the flat side of a cubic zirconia solid immersion lens (SIL) with a refractive index of $n = 2.15$. The SIL was thermally contacted to a cold finger in a liquid helium flow cryostat. A continuous-wave Ti:saphhire laser operating at 976~nm was locked to a home-built high-finesse cavity to achieve a spectral linewidth $<10$~kHz and long-term spectral stability. For addressing the $^3H_4$-$^3P_0$ transition, the resulting laser light was frequency doubled to 488~nm in a resonant cavity and could be scanned over 600~MHz using an acousto-optical modulator (AOM) in double-pass mode. Broader spectral investigations could be obtained by stitching these scans. The laser beam was then coupled to a microscope objective (numerical aperture 0.75) that was scanned by a three-dimensional piezoelectric actuator. We used a single-photon detector with very low dark counts  to measure the fluorescence signal from the sample in reflection and employed another avalanche photodiode (APD) to detect the laser light in transmission through a bore in the cold finger.

The first step of our measurements consisted of imaging individual YSO crystallites on the SIL surface. Here, we monitored the reflected excitation light on a photodetector to record the interference between the light fields scattered by the small crystals and the SIL surface.  Figure~\ref{setup}D displays a scanning image of the crystallites using this interferometric scattering (iSCAT) contrast, which depends on the size and index of refraction of the object under illumination~\cite{lindfors04detection}. For the experiments discussed below, we used the nanocrystal in the middle of the image, which we estimate from the cross section in the inset of Fig.~\ref{setup}D to be of the order of 100-200 nm in size. Considering the weak signal of an ion, even small sample drifts can complicate the measurements. We, thus, used the iSCAT signal to also actively lock the laser spot to the nanocrystal, making it possible to study a single ion for hours. To avoid population trapping in the hyperfine levels of the ground state, we used a second AOM to generate two frequencies at $f_1=f_2-10.19$~MHz and $f_3=f_2+17.3$~MHz in addition to the laser frequency ($f_2$). This allowed us to simultaneously excite all three hyperfine levels of the ground state.  

Figure~\ref{single_ion}A shows an example of an excitation fluorescence spectrum when frequencies $f_1$, $f_2$ and $f_3$ were scanned over 4~GHz. To reduce the measurement time, this spectrum was recorded with a laser linewidth of about 1 MHz without using cavity locking. A background fluorescence as low as about 20-30 counts per second allows the detection of narrow spectral features. Figure~\ref{single_ion}B plots a zoom into three narrow resonances recorded with the narrow-band laser ($<$ 10 kHz). In most cases, such peaks were spectrally stable over measurement times of the order of hours although we also observed some spectral diffusion. The sparse appearance of the observed narrow spectral lines lets us attribute them to the emission of single Pr$^{3+}$ ions~\cite{Moerner:89}. 

Each of the spectrally detected ions can be spatially mapped in fluorescence. Figure~\ref{single_ion}C presents a laser scanning image of the fluorescence signal when the laser frequency was tuned to a narrow resonance. Fitting the image of the diffraction-limited spot lets us localize the ion with a precision of 5 nm. By repeating this procedure for each resonance, one can co-localize all the ions in the sample beyond the diffraction limit. Figure~\ref{single_ion}D illustrates this idea for the three ions of Fig.~\ref{single_ion}B. Combination of this type of super-resolution microscopy and high-resolution spectral data will open the way for quantitative studies of near-field coupling among several ions~\cite{Hettich:02}.

\begin{figure*}
\includegraphics[width=0.8\textwidth]{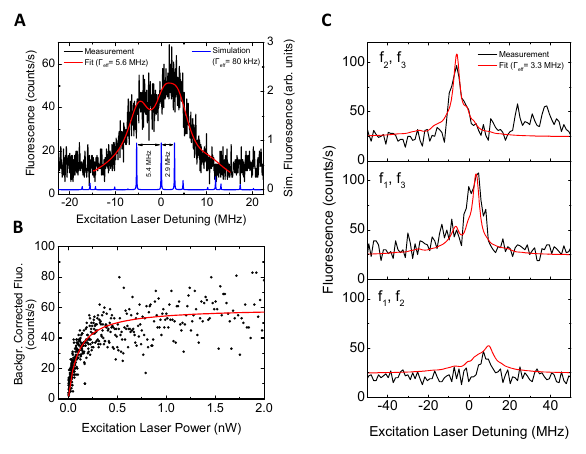}
\caption{\label{high_resolution}
\textbf{A}, The blue spectrum depicts a simulated spectrum, assuming a transition linewidth of 80 kHz. The black curve plots an experimental high-resolution excitation spectrum of a single ion. The substructure originating from the excited state hyperfine splitting is clearly visible. The red curve displays a fit, assuming a full width at half-maximum of 5.6 MHz. \textbf{B}, Background-corrected fluorescence of a single ion as a function of the excitation power. The red curve denotes a fit assuming a three-level system. \textbf{C}, Excitation spectra of a single ion illuminated by different combinations of only two laser frequencies. The legend in each plot denotes the frequencies that were on. The red curves were obtained by fitting all three two-frequency spectra simultaneously.}
\end{figure*}

Up to this point, we treated the excited state as one level. However, one has to bear in mind that the $^3P_0$ state consists of three hyperfine levels at separations of 5.4 and 2.9 MHz, which become resonant with the three laser frequencies $f_1$, $f_2$ and $f_3$ at different detunings. The blue spectrum in Fig.~\ref{high_resolution}A plots the fluorescence signal that is expected from coupling the three ground state hyperfine levels to the three excited state hyperfine levels if one assumes a linewidth of 82~kHz. The three larger peaks occur whenever the laser detuning matches the transition of all ground state hyperfine levels to one of the excited state hyperfine levels. The intensities of the other peaks are smaller because optical pumping sets in if not all ground state levels are addressed simultaneously. The displayed intensities were calculated using a simple rate equation model, where the ion was described as a six-level system. 

The spectra in Fig.~\ref{single_ion}B do not mirror the expected multitude of narrow resonances in Fig.~\ref{high_resolution}A, but spectra recorded at higher scan-pixel resolution (see black curve in Fig.~\ref{high_resolution}A) clearly show a substructure. Having confirmed that the observed linewidths of about 12 MHz were not limited by the sample temperature (see Supplementary Materials), we attribute this broadening to spectral diffusion caused by the strain in the milled crystallites. The red curve shows that the measured spectrum in Fig.~\ref{high_resolution}A can be matched if we consider the individual transitions to be broadened to 5.3 MHz. We point out that although size-related deviations of the linewidth are expected in very small nanocrystals~\cite{Meltzer:00}, there is no fundamental reason for the existence of dephasing in crystallites of the order of 100 nm or larger. 

Figure~\ref{high_resolution}B plots the fluorescence signal at the peak of a recorded resonance as a function of the excitation power. We find a measured saturation power of 98 pW in each of the three laser beams equivalent to an intensity of $I_{sat} \simeq 46~\frac{mW}{cm^2}$. The measurement also indicates a maximum detected count rate of about 60 photons/s, corresponding to an emission rate of 700-1000 photons/s when accounting for our measured detection efficiency of 11\% and estimated collection efficiency of about 54-78\% through the SIL (see Supplementary Materials). This is about 6-8 times smaller than the emission signal of about 5800 photons/s expected from rate equations, when taking into account the upper, intermediate and ground states $^3P_0$, $^1D_2$ and $^3H_4$, respectively (see Supplementary Materials). We attribute the source of this discrepancy to long-lived states at energies below the $^1D_2$ state (see Fig.~\ref{setup}A), which could act as bottle neck. Although the branching ratios of decay into these states can be moderate (see Fig. S1A), a lifetime of the order of 500 $\mu s$, which is typical for 4f-shell transitions, would suffice to rectify the observed discrepancy. It should be noted that proper optical pumping out of these trap states would allow one to recover the maximum fluorescence rate.

Earlier we argued that we needed three laser frequencies to prevent shelving in the upper hyperfine levels of the ground state. While Fig. S1E in the Supplementary Materials verifies this statement, the black spectra in Fig.~\ref{high_resolution}C illustrate that the fluorescence survives if one turns off one of the frequency components $f_1$, $f_2$ or $f_3$. The fits presented by the red solid curves in Fig.~\ref{high_resolution}C indicate that the recorded data can be reproduced if a homogeneous linewidth of 3.3 MHz is assumed for this ion (different from the one examined in Fig.~\ref{high_resolution}A). Theoretical simulations show that the line shape and peak heights of each combination depend on the homogeneous linewidth (see Supplementary Materials).

\begin{figure*}
\includegraphics[width=0.8\textwidth]{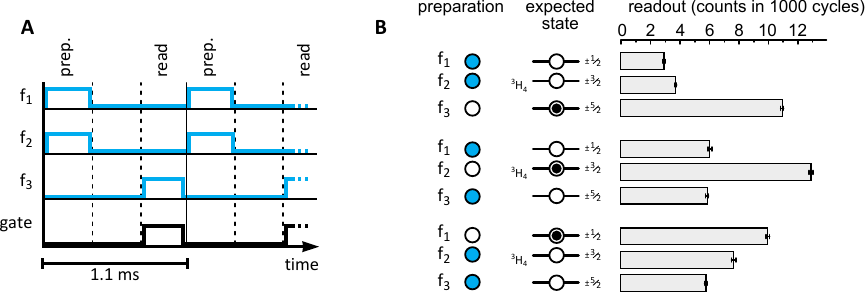}
\caption{\label{pulsing}
\textbf{A}, An example of a pulse sequence for state preparation and read-out of a single ion. The ion is pumped into one of the hyperfine levels by two laser frequencies that are resonant with the other two hyperfine levels. A single laser frequency is used for the read out. \textbf{B}, Three different scenarios for preparing the ion in the lowest, middle, or upper hyperfine levels of the ground state. In all three cases, the read-out clearly shows a high fluorescence yield only when the laser is resonant with the previously populated hyperfine level. Each sequence was cycled for 100~s to obtain a sufficient signal-to-noise ratio. }
\end{figure*}

The above measurements illustrate that a single-ion can be prepared in a selected ground state by continuous optical pumping of the population to the desired state. To demonstrate the read out of the resulting quantum states, we implemented a pulsed excitation scheme sketched in Fig.~\ref{pulsing}A. The pulses were generated by a digital delay and a pulse generator and were applied independently to the three laser lines by the same AOM that produced $f_1$, $f_2$ and $f_3$. Two laser frequencies were turned on to pump the population out of two hyperfine levels to the third one. The pump duration of $344~\mu s$ was chosen to reach a population transfer of about 90\%. A delay of $378~\mu s$ was implemented to allow the ion to decay from the excited state before applying the read-out pulse of $378~\mu s$ at the frequency of the third hyperfine level. This pulse drove the population for a few cycles until it was again trapped in the other two levels. We applied a gate pulse during the read out to discriminate against background counts. Figure~\ref{pulsing}B displays the outcome of three sets of such measurements over 100~s. It is clear that if the read-out pulse is resonant with the state in which the ion was prepared, the fluorescence signal is maximized. These data exhibit the ease with which a single rare-earth ion can be optically prepared in a hyperfine level of the electronic ground state although in the current investigation the efficiency was compromised due to the broadening of the transitions.

High-resolution spectroscopy of rare earths at the single ion level combined with microscopy and magnetic resonance spectroscopy~\cite{fraval04method, Heinze:11, Heinze13minutememory} would provide unprecedented opportunities for exploring fundamental phenomena such as dipole coupling, energy transfer and phonon broadening in materials of technological relevance~\cite{Jacquier} without the ambiguities associated with ensemble inhomogeneities. Furthermore, our results stimulate new experimental frontiers for quantum information science~\cite{kimble08quantum, Awschalom:13}. In particular, on-chip nanoscopic waveguides and interferometers~\cite{OBrien:07} made of Pr:YSO could offer an ideal platform for connecting atom-like qubits via efficient photonic channels. The only shortcoming of the rare earth systems, namely their weak emission that led many groups to consider their direct detection to be unfeasible via the 4f-shell transitions~\cite{kolesov12optical}, can be readily improved in future works. First, higher collection~\cite{lee11} and detection efficiencies could increase the fluorescence count rates to the $10^4$ counts/s level. Moreover, the spontaneous emission rate and branching ratios can be modified by hundreds to thousands times via application of microcavities~\cite{mcauslan09strong-coupling} and nano-antennas~\cite{Chen:12}. We hope that our work inspires further activities towards the detection of single ions also in other guest-host systems. 

This project was supported by a European Research Council advanced grant (SINGLEION) and the Max Planck Society. We also acknowledge seed funding from ETH Zurich in its initial stage before our laboratory moved to Erlangen.

\newpage

\section{Supplementary Material}

\bigskip

\textbf{Detection, spectroscopy and state preparation of a single praseodymium ion in a crystal}

\textit{T. Utikal, E. Eichhammer, L. Petersen, A. Renn, S. G\"otzinger, V. Sandoghdar}

\bigskip
\bigskip
\setcounter{figure}{0}
\renewcommand{\thefigure}{S\arabic{figure}} 

\noindent \textbf{Fluorescence spectroscopy}

\begin{figure*}
\includegraphics[width=0.85\textwidth]{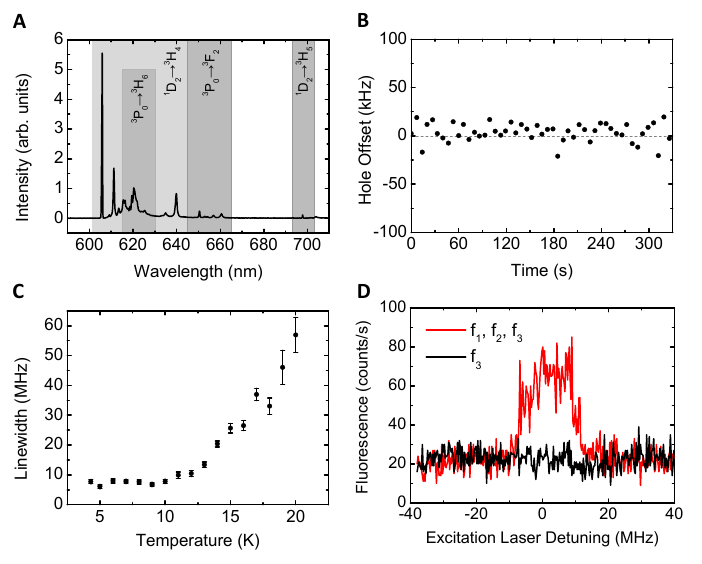}
\caption{\label{supp_1}
\textbf{A}, Fluorescence emission spectrum of bulk Pr:YSO with excitation light polarized parallel to the $D_1$-axis measured at T = 4.3~K. \textbf{B}, Position of the hole center after a single burn pulse and repeated scans. The measurement confirms the long-term frequency stability. \textbf{C}, Temperature dependence of the linewidth of a single ion resonance. \textbf{D}, Fluorescence signal recorded when addressing all hyperfine levels of the ground state (red) compared to the case where only the lower level was excited (black). In the latter case, population is trapped in the higher hyperfine levels.}
\end{figure*}

Figure~\ref{supp_1}A shows a fluorescence spectrum recorded from bulk Pr:YSO. Here we set the polarization of the incident light parallel to the $D_1$ axis of the crystal and excited the ions at the center of the inhomogeneous broadening at T = 4.3~K. The main peak at 606~nm corresponds to the transition from $^1D_2$ to $^3H_4$. The spectral features at longer wavelengths correspond to transitions from either $^3P_0$ or $^1D_2$ to other intermediate levels as indicated.

\bigskip

\bigskip
\noindent \textbf{Spectral hole burning}

For the spectral hole burning experiments we replaced the microscope objective by a lens with focal length f=300~mm in order to interact with a multitude of ions under a uniform illumination. The excitation light was re-collimated behind the cryostat and sent to the detector. We tuned the laser to the center of the inhomogeneous broadening and burnt a spectral hole with a short but intense laser pulse using only a single laser frequency. To interrogate the spectral hole, the laser intensity was strongly reduced and the frequency was scanned over the hole while recording the transmitted signal. Before commencing the next cycle, the hole was cleared by heating the sample to 12~K and subsequently cooling to 4.3~K. The hole burning spectrum in Figure 1B of the main article is an average of 15 scans. The burn pulse had about $5~\mu$W for 100~ms, and the scan was performed with 25~nW over a scan range of 60 MHz within 20~s. 

The strong burn fluence led to the power broadening of the center hole. To get around this, we repeated the experiments with a burn pulse of only 160~nW for 200~ms and subsequently scanned the laser with 0.7~nW in 500~ms over 4~MHz. The inset of Fig. 1B in the main text shows an average of 20 scans of the center hole measured at low fluence. A Lorentzian fit to the hole yields a FWHM of 172~kHz. Since a laser beam with linewidth $\Gamma_{\textnormal{laser}}$ interacts with transitions of linewidth $\Gamma_{\textnormal{hom}}$ during the burn and scan sequence, the total hole width is given by $\Gamma_{\textnormal{hole}} = 2(\Gamma_{\textnormal{laser}}+\Gamma_{\textnormal{hom}})$. Accounting for the homogeneous linewidth of $\Gamma_{\textnormal{hom}}$=82~kHz deduced from lifetime measurements, we estimate a laser linewidth of ($4\pm2.5$)~kHz. 

In order to estimate the long-term laser frequency stability, we burnt a hole with a high fluence to achieve a good signal-to-noise ratio and repeatedly scanned the hole for several minutes (without heat and re-cool cycles). Each scan was fitted with a Lorentzian to determine the hole center frequency as a function of time. Long-term drifts of the laser frequency would lead to a shift of the hole position while dephasing in the sample would result in a broadening and reduction of the hole depth. Figure~\ref{supp_1}B displays the hole center distribution over six minutes in 7~s intervals. The fit uncertainty of the hole centers is 6~kHz. Together with the standard deviation of the data points (9~kHz) we can estimate the frequency stability to be $\sqrt{(9~\textnormal{kHz})^2-(6~\textnormal{kHz})^2}\approx 7~\textnormal{kHz}$ on a time scale of 7~s. A further drift on the scale of minutes is not observable. This is also confirmed by the spectral stability of the single ion resonances in consecutive scans.
\bigskip

\noindent \textbf{Temperature dependence}

The single ion resonances in the excitation spectra are broader than expected. Figure~\ref{supp_1}C shows the temperature dependence of the measured linewidths. Here we plotted the full width at half-maximum of an observed resonance (see Fig. 3A of the main manuscript), neglecting the underlying structure of the three hyperfine transitions. We find that as expected, the linewidth decreases at lower temperatures, but it levels off at about 10~K. We also emphasize that the line form and substructure did not change down to a temperature of 4.3 K. We conclude that the sample was sufficiently cooled and the linewidth was not affected by temperature broadening. 
\bigskip

\noindent \textbf{Simulation of the excitation spectra}

For a better understanding of the single ion fluorescence, we set up a rate equation model, where the ion is described as a six-level system, taking into account the three ground and three excited hyperfine states, neglecting the intermediate state $^1D_2$. The values of the hyperfine splittings are taken from hole burning measurements and no population transfer within the manifolds of the excited or ground states is allowed. The transfer from the ground to the excited state is driven by the laser frequencies with the splittings of 10.19~MHz and 17.3~MHz which are scanned synchronously. In this simple model, each transition is taken to be Lorentzian. The blue lines in Fig. 3A of the main text plot the obtained fluorescence rate as a function of the laser frequency detuning for a three-frequency excitation if the linewidth of the transitions is assumed to be 82~kHz. The spectrum is dominated by three peaks separated by the splittings of the excited state hyperfine levels. These peaks occur whenever the laser detuning matches the transition of all ground states to one of the exited state hyperfine levels. They are not affected by population trapping in one of the ground states. This is different for the other peaks, which are reduced in amplitude since optical pumping sets in if not all ground states are addressed simultaneously.

Exciting a single ion with a single laser frequency results in a strong reduction of the fluorescence signal since the ion is quickly pumped to the other two hyperfine levels of the ground state, where it is not available for further excitation. The measurement in Fig.~\ref{supp_1}D confirms this by comparing the signals when all three frequencies $f_1$, $f_2$, $f_3$ were on with the case where only $f_3$ was on. 

\begin{figure*}
\includegraphics[width=0.9\textwidth]{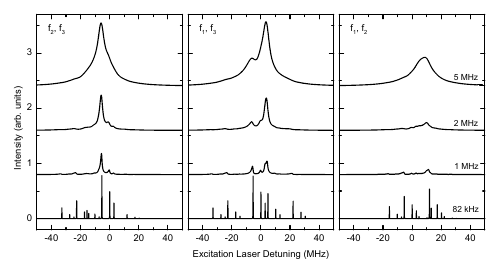}
\caption{\label{supp_2} The three panels show the simulated spectra for the combinations of $f_2$ and $f_3$, $f_1$ and  $f_3$, $f_1$ and $f_2$. In each panel the spectra for four choices of linewidths are displayed.}
\end{figure*}

We showed in the main article (Fig. 3C) that we do observe a fluorescence signal when two of the three frequencies are on. We also showed that by using the above-mentioned simple model, we could fit the three spectra simultaneously to deduce a homogeneous linewidth of 3.3 MHz. In Fig.~\ref{supp_2}, we study the dependence of the spectrum on the linewidth of the transitions. The three panels present the simulated spectra for the combinations of $f_2$ and $f_3$, $f_1$ and  $f_3$, $f_1$ and $f_2$. In each panel we plot the spectra for linewidths 82 kHz, 1 MHz, 2 MHz and 5 MHz. It is clear that the choice of linewidth not only determines the spectral line shape but also the signal height. 

\bigskip

\noindent \textbf{Maximum emission rate at saturation}

To model the saturation behavior of an ion, we first assume a rate-equation model involving the ground state $^3H_4$, the excited state $^3P_0$, and the intermediate state $^1D_2$. The decay from $^3P_0$ to $^1D_2$ and $^3H_4$ is modeled according to the branching ratios, which were determined from fluorescence spectra (such as Fig.~\ref{supp_1}A) to be $^3P_0\rightarrow{}^1D_2 = 39\%$. The decay rates of $^3P_0$ and $^1D_2$ were chosen according to the experimentally measured values of $1.95~\mu$s and $166~\mu$s, respectively. This simple model predicts a photon rate of about 5800 photons/s emerging from the $^1D_2\rightarrow{}^3P_0$ transition at saturation.

To compare this value with the detected count rate of 60 photons/s at saturation (Fig. 3B of the main article), we have to take into account the detection and collection efficiencies of our system. The former was measured to be 11\% by examining the transmission of a laser beam at the emission wavelength. However, precise measurement of the collection efficiency is more tedious. Instead, it can be estimated from calculations (e.g. according to J.-Y. Courtois et al., Phys. Rev. A 53, 1862 (1996)). Here we assume the emitter to be placed at a distance of 50~nm from the flat side of the solid immersion lens (SIL) at a refractive index contrast of $n_{\textnormal{SIL}}/n_{\textnormal{air}} = 2.15$. We obtain the fraction of photons that is within the collection solid angle to be 78\% for a dipole moment parallel to the substrate and 54\% for a dipole moment normal to the SIL interface. Hence, we deduce the maximum emission rate to be 700-1000 photons/s, which is smaller than the above-mentioned estimate of 5800.  

We believe the reason for the observed lower emission rate is decay into the states such as $^3H_6$, $^3H_5$, $^3F_{2,3,4}$ lying above $^3H_4$. Although the branching ratio of the fluorescence decay into these states is not large, long lifetimes (Kolesov et al., Nature Comm. 3, 1029 (2012)) can lead to population trapping and lower emission at saturation. To model this, we introduced a fourth state in the rate equations. Using a more differentiated account of the recorded emission spectra in a bulk crystal, we determined the branching ratios to be $^3P_0\rightarrow{}^1D_2 = 39\%$, $^3P_0\rightarrow{}^3H_4 = 13\%$, and $^3P_0\rightarrow{}^3H_6 / ^3H_5 / ^3F_{2,3,4} = 48\%$. All decays from the intermediate levels were assumed to be invisible to the detection. The lower measured count rates can be explained if one assumes a lifetime of the order of $500~\mu$s for the intermediate state, which is a realistic order of magnitude for rare earth ions (R. M. Macfarlane, J. Luminesc. 100, 1 (2002) and Ref. 4 of the main article).

\end{document}